**Unlocking the potential of polymer networks for high strength through topology**

Zheqi Chen[1, 2. †], Yu Zhou[1, †], Chenghai Li[1], Ho Cheung Shum[2,3], Zhigang Suo[1,*]

[1] John A. Paulson School of Engineering and Applied Sciences, Harvard University, MA 02138
[2] Advanced Biomedical Instrumentation Centre, Hong Kong Science Park, HKSAR
[3] Department of Mechanical Engineering, The University of Hong Kong, HKSAR

[†] These authors equally contributed to this work.
[*] To whom correspondence should be addressed. Email: suo@seas.harvard.edu (Z.S.)

## Abstract

A covalent polymer network is orders of magnitude weaker than individual covalent bonds. This large gap reveals an unrealized potential. The weakness of a network comes not from individual covalent bonds, but from how they are connected. Here we unlock this potential by identifying a topological criterion, involving optimally sparse crosslinks, long chains, and dense entanglements. Optimizing each topological variable alone is insufficient for high strength, yet exceptional strength emerges when they are combined within a narrow regime. We describe the physics of this synergy, and use the criterion to strengthen networks widely used but known to be weak. For example, we prepare a polyacrylamide hydrogel of 90 wt% water, amplifying strength from a commonly reported value of ~0.1 MPa to ~6.5 MPa. The exceptional strength is achieved without sacrificing other properties, such as softness, toughness, and elasticity. The criterion establishes topology as a general design principle for strong polymer networks.

## Main

Many applications rely on rubbery polymer networks for their elasticity and strength. However, elasticity and strength are affected by the network architecture differently. In a rubbery polymer, covalent bonds link monomers into chains, and crosslink the chains into a network. The covalent network enables large and elastic deformation, but is orders of magnitude weaker than an individual covalent bond[1-5]. The large gap in strength reveals an unrealized potential. Despite numerous efforts[6-11], ample room exists to elevate the strength of a network.

The weakness of a network comes not from individual covalent bonds, but from how they are connected. The topology of a network is characterized by linelike features (chains, strands, and dangling ends) and junctionlike features (crosslinks and entanglements) (Fig. 1a). We identify a criterion for the topology of strong networks, and apply it to strengthen networks previously known to be weak. For example, a polyacrylamide hydrogel, containing ~90 wt% of water, typically has a strength of ~0.1 MPa [11-20], which we now amplify by nearly two orders of magnitude, to ~6.5 MPa (Fig. 1b, Supplementary Video 1). The criterion captures how the topological features interplay to affect strength in several ways. First, optimally sparse crosslinks maximize strength; overly dense crosslinks result in concentration of high tension to a small fraction of covalent bonds, whereas overly sparse crosslinks result in a large fraction of dangling ends (Fig. 1c). Second, long chains elevate the peak strength by reducing the lower limit of the crosslink density. Third, dense entanglements resist excessive deformation without concentrating tension. We validate the criterion in gels and elastomers.

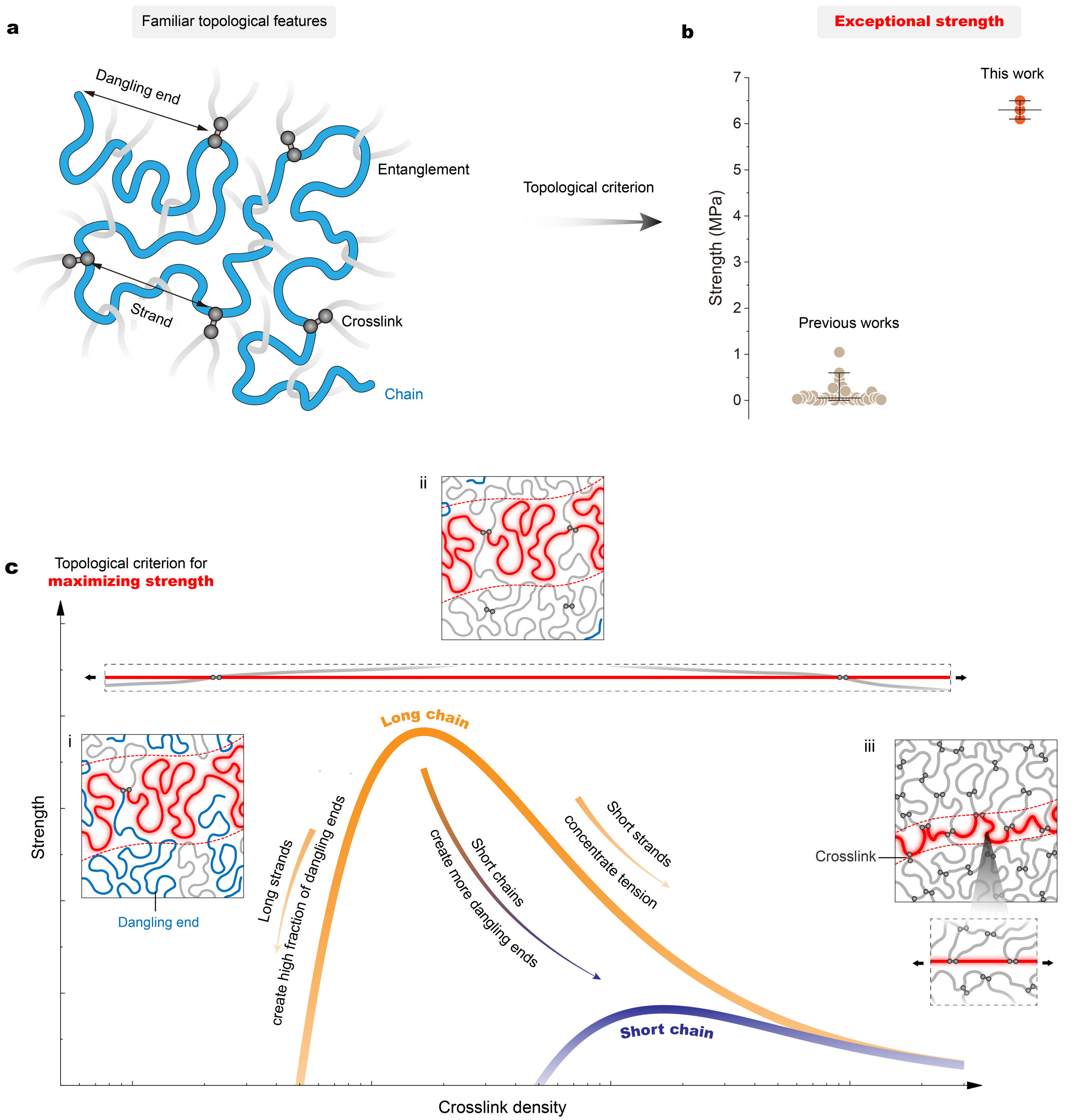

**Fig. 1. Unlocking the potential of polymer networks for strength through the design and synthesis of network topology.** (a) The topology of a network is characterized by various features. (b) A topological criterion that elevates the strengths of polyacrylamide hydrogels by nearly two orders of magnitude. The data of previous works are collected from literature[1,11-18]. (c) The topology of a network affects its strength. The curves are drawn by using equation (3).

## Results

### Physical origin of the topological criterion

We describe the topology of a network by five features: chains, strands, dangling ends, crosslinks, and entanglements (Fig. 1a). Once the numbers of three features are fixed, so are the numbers of the other two. For example, let $N$, $N_{\mathrm{c}}$, and $N_{\mathrm{e}}$ be the numbers of chains, crosslinks, and entanglements in a network. Each chain has two dangling ends, so that the network has $2N$ dangling ends. We synthesize the network using bifunctional crosslinkers, so that each crosslink connects four segments. Of the $4N_{\mathrm{c}}$ segments, $2N$ are dangling ends, and $4N_{\mathrm{c}} - 2N$ are strands. Each strand is connected to two crosslinks, so that the network has $2N_{\mathrm{c}} - N$ strands.

The topological features affect strength collectively. We describe the physical origin of this collective action using an intermediate quantity: the effective strength. Define the nominal stress $s$ by the force applied to a sample divided by the cross-sectional area of the undeformed sample, the stretch $\lambda$ by the length of the deformed sample divided by the length of the undeformed sample, and the true stress $\sigma$ by the applied force divided by the cross-sectional area of the deformed sample. During deformation, a network changes shape but not volume, so that $\sigma = s\lambda$. Because hydrogels are commonly used in water, we submerge the as-synthesized hydrogels in pure water to swell to equilibrium, and then measure their properties. Define the polymer content $\varphi$ by the volume of dry polymer divided by the volume of the fully swollen gel. Water in a hydrogel does not bear load. Following our previous work[9], we define the effective stress by

$$\Sigma = s\lambda/\varphi \tag{1}$$

For a hydrogel, $\varphi < 1$. For an elastomer, $\varphi = 1$.

The nominal stress and the stretch at rupture define the nominal strength $s_{\mathrm{f}}$ and stretchability $\lambda_{\mathrm{f}}$. The effective strength, $\Sigma_{\mathrm{f}} = s_{\mathrm{f}}\lambda_{\mathrm{f}}/\varphi$, is computed using the measured nominal strength, stretchability, and polymer content. The nominal strength $s_{\mathrm{f}}$ quantifies the force that can be sustained by a material, whereas the effective strength $\Sigma_{\mathrm{f}}$ quantifies the capability of the network to recruit covalent bonds to bear high tension. The former is important in applications, and the latter aids in molecular understanding. By definition, the two measures of strength are related by

$$s_{\mathrm{f}} = \Sigma_{\mathrm{f}}\varphi/\lambda_{\mathrm{f}} \tag{2}$$

To achieve a high nominal strength, a network should recruit a large fraction of covalent bonds to bear high tension, maintain high polymer content, and avoid excessive stretchability. We next relate these physical considerations for strength to the topological features of a network.

We have characterized the topology of a network using $N$, $N_{\mathrm{c}}$, and $N_{\mathrm{e}}$ as the three independent variables. In general, the strength of a network is a function of the three variables. Here we propose the following strength-topology relation:

$$s_{\mathrm{f}} = N_{\mathrm{c}}^{-2/3} \frac{2N_{\mathrm{c}}-N}{2N_{\mathrm{c}}+N} f(N_{\mathrm{e}}) \tag{3}$$

The first factor comes from a scaling relation between the effective strength and crosslink density (Supplementary Note 1),

$$\Sigma_{\mathrm{f}} \sim N_{\mathrm{c}}^{-2/3} \tag{4}$$

This scaling relation approximately fits experimental data, and is rationalized as follows. The tension in an individual strand is low when it is coiled, and rises steeply only as it approaches full extension. When stretched, a network forms and breaks high-tension paths in succession (Fig. 1c, iii)[21-24]. Each path consists of a sequence of crosslinked strands from one end of the network to the other. The paths concentrate high tension while the surrounding strands remain coiled and carry low tension. Reducing crosslinks recruit more repeat units in bearing high tension (Fig. 1c, ii). Furthermore, crosslinks and entanglements behave differently in distributing high tension. Crosslinks do not slip, and concentrate high tension to a small fraction of covalent bonds. By contrast, entanglements slip, and homogenize tension among strands[8,9,15,25-27]. When a network is stretched, entanglements act as pulleys, and do not impede the high tension from transmitting over long strands. Consequently, entanglements do not change the number of repeat units for bearing high tension.

In our previous work, the scaling relation $\Sigma_{\mathrm{f}} \sim N_{\mathrm{c}}^{-2/3}$ does not fit data for networks with overly sparse crosslinks.[9] Here we introduce the second factor, $(2N_{\mathrm{c}} - N)/(2N_{\mathrm{c}} + N)$, to account for the fraction of strands. As noted above, a network of $N$ chains and $N_{\mathrm{c}}$ crosslinks has $2N$ dangling ends and $2N_{\mathrm{c}} - N$ strands. When a network is stretched, strands bear tension, but dangling ends do not. Consequently, the effective strength scales as

$$\Sigma_{\mathrm{f}} \sim N_{\mathrm{c}}^{-2/3} \frac{2N_{\mathrm{c}}-N}{2N_{\mathrm{c}}+N} \tag{5}$$

When $N_{\mathrm{c}} >> N$, the factor $(2N_{\mathrm{c}} - N)/(2N_{\mathrm{c}} + N) \to 1$, and the fraction of dangling ends is negligible. When $N_{\mathrm{c}} = N/2$, the factor is zero, below which the network does not form.

Because entanglements restrain swelling and stretchability, the ratio $\varphi/\lambda_{\mathrm{f}}$ is an increasing function of entanglement density. For a network in which entanglements greatly outnumber crosslinks, $N_{\mathrm{e}} \gg N_{\mathrm{c}}$, the polymer content and stretchability are set by the entanglement density, and are insensitive to the crosslink density[15]. Here we assume that the ratio $\varphi/\lambda_{\mathrm{f}}$ varies with $N_{\mathrm{e}}$, but is insensitive to $N_{\mathrm{c}}$ and $N$. The factor $f(N_{\mathrm{e}})$ is a monotonically increasing function of $N_{\mathrm{e}}$.

These considerations together lead to the strength-topology relation (3). We should note that this relation is approximate, but we will show that the relation describes the trends of our experimental data. We have chosen $N$, $N_{\mathrm{c}}$, and $N_{\mathrm{e}}$ as the three independent variables. We next describe how each affects strength when the other two are fixed. First, for fixed $N$ and $N_{\mathrm{e}}$, strength varies non-monotonically with $N_{\mathrm{c}}$, and peaks at an intermediate number of crosslinks (Fig. 1c, ii). Too few crosslinks result in a high fraction of dangling ends (Fig. 1c, i), whereas too many crosslinks concentrate tension (Fig. 1c, iii). Second, for fixed $N_{\mathrm{c}}$ and $N_{\mathrm{e}}$, strength monotonically increases with chain length; the longer the chains, the lower the fraction of the dangling ends. Third, for fixed $N$ and $N_c$, strength increases with $N_e$. Entanglements restrain stretchability and swelling.

The strength-topology relation (3) states the criterion for strong networks in terms of three independent topological features: optimally sparse crosslinks, long chains, and dense entanglements. The three topological features act synergistically. Sparse crosslinks promote tension spreading but impose penalties—excessive fraction of dangling ends, stretchability, and swelling. Long chains restore connectivity: the longer the chains, the fewer the crosslinks needed to form a network. Dense entanglements suppress excessive stretchability and swelling. Long chains and dense entanglements thus remove the penalties of sparse crosslinks while preserving the ability of a network to spread high tension.

**Experimental validation of the criterion**

We ascertain the topological criterion by experiments. We choose the polyacrylamide (PAAm) hydrogel as a model system for two reasons. First, PAAm hydrogels are widely used but known to be weak, with a strength typically at ~0.1 MPa [11-18]. Second, the three topological features can be changed by precursor compositions.

The hydrogels are synthesized from precursors of four components: monomer, crosslinker, photoinitiator, and water (Fig. 1a). The composition of a precursor is defined by three molar ratios: initiator-to-monomer ratio $I$, crosslinker-to-monomer ratio $C$, and water-to-monomer ratio $W$. The polymerization is triggered under an ultraviolet light of a fixed intensity. At the fixed $W$ and fixed light intensity, the amount of initiator controls the radical concentration during polymerization, and therefore determines the polymer chain length: increasing $I$ produces more chains, and shortens each chain (Supplementary Note 2). The amount of crosslinker determines the strand length: increasing $C$ produces more crosslinks, and each strand is shorter. To create dense entanglements, we fix $W = 2$. A precursor of such an unusually low $W$ results in crowded polymers, which densely entangle[15].

To test the topological criterion, we prepare two series of hydrogels, one from $I = 10^{-3}$, and the other from $I = 10^{-4}$. The chain length of the former is shorter than the latter, as confirmed by the lower viscosity of the corresponding solutions of uncrosslinked PAAm prepared under the same polymerization conditions (Supplementary Fig. 1). We therefore name the former as short-chain hydrogels, the latter as long-chain hydrogels. Within each series, $C$ is varied to change the crosslink density. As noted before, after polymerization, each as-prepared hydrogel is immersed in pure water to swell to equilibrium. We test all the properties at the fully swollen state.

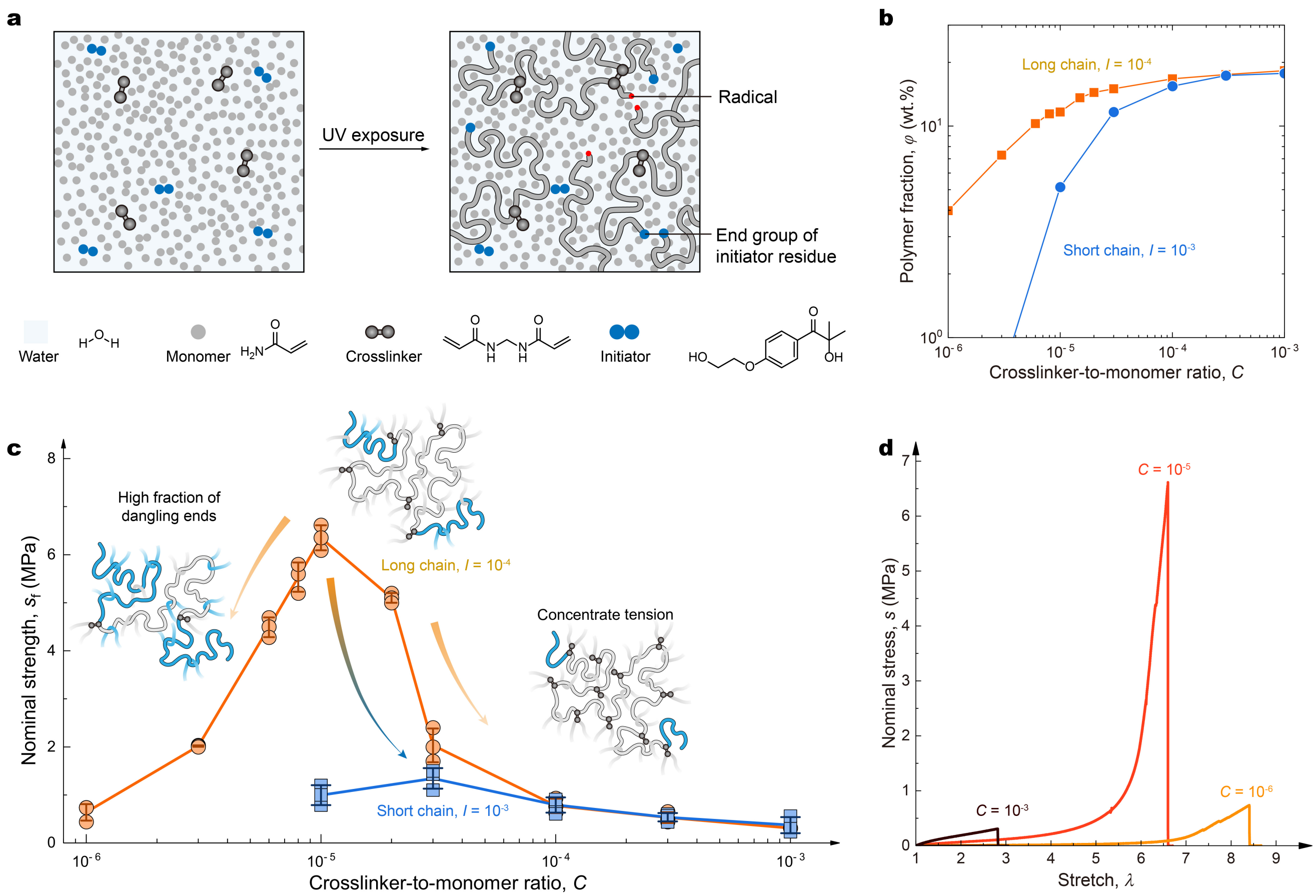

**Fig. 2. Effect of crosslink density and chain length on strength.** (a) The hydrogel is polymerized from a precursor of four components. (b) Polymer fraction as a function of $C$ at $I = 10^{-3}$ and $I = 10^{-4}$. (c) Nominal strength as a function of $C$ at $I = 10^{-3}$ and $I = 10^{-4}$. Data are presented as mean values ± standard deviations (s.d.) of $n = 3$ replicates. (d) Nominal stress–stretch curves for hydrogels with three representative values of $C$ at $I = 10^{-4}$.

Long chains extend the lower limit of crosslink density for maintaining a network (Fig. 2b). At $C > 10^{-4}$, the two series of hydrogels have comparable polymer fraction. At $C < 3\times10^{-5}$, the short-chain hydrogels swell substantially more than long-chain hydrogels. This divergence is consistent with the formation of a large fraction of dangling ends when short chains are sparsely crosslinked. Similar trends are observed by the storage modulus and mechanical loss tangent as a function of crosslink density (Supplementary Fig. 2).

For both series of hydrogels, the strength peaks at an intermediate crosslink density, but the peak strength and optimal value of $C$ differ between the two series (Fig. 2c). The peak strength is ~6.5 MPa at $C = 10^{-5}$ for the long-chain hydrogel, and is ~ 1 MPa at $C = 3\times10^{-5}$ for the short-chain hydrogel. These results provide a direct experimental validation of Fig. 1c, and the strength-topology relation (3).

As remarked before, the non-monotonic dependence of strength on crosslink density reflects two competing effects of crosslinks (Fig. 2d). Excessively dense crosslinks concentrate tension to a small fraction of strands, so that the hydrogel is weak. Excessively sparse crosslinks create a large fraction of dangling ends, so that the hydrogel is also weak. For a given chain length, the strength peaks at an intermediate crosslink density, where the two competing effects are balanced.

It is the long chain that enables the exceptional strength of 6.5 MPa. Increasing chain length shifts the optimum towards lower crosslink density and amplifies the peak strength. At $C = 10^{-5}$, the long-chain hydrogel enables long strands while maintaining connectivity, allowing the high tension to spread to many repeat units. By contrast, the short-chain hydrogel loses connectivity at $C = 10^{-5}$ because of the excessive fraction of dangling ends, and strength decreases.

**Synthesis sets additional boundaries**

Synthesis sets additional boundaries in which topologies are experimentally accessible. It is expected that a lower $I$ would generate longer chains, which should further enhance strength. Experimentally, however, this trend breaks down at an extremely low fraction of the initiator (Fig. 3). We fix $C = 10^{-5}$, and further reduce $I$. The temperature of the reaction cell is recorded as a function of time (Fig. 3a, Supplementary Fig. 3). At all investigated values of $I$, polymerization is captured by a characteristic temperature peak. As $I$ decreases, the peak position shifts to longer reaction times, consistent with slower polymerization kinetics. As $I$ decreases, the polymer fraction of the fully swollen hydrogel increases and plateaus, as expected (Fig. 3b). These results confirm that lower $I$ indeed promotes the formation of longer chains. However, contrary to the expectation, the strength does not continuously increase as $I$ decrease. Instead, strength peaks at $I = 10^{-4}$, and reduces sharply at extremely low $I$.

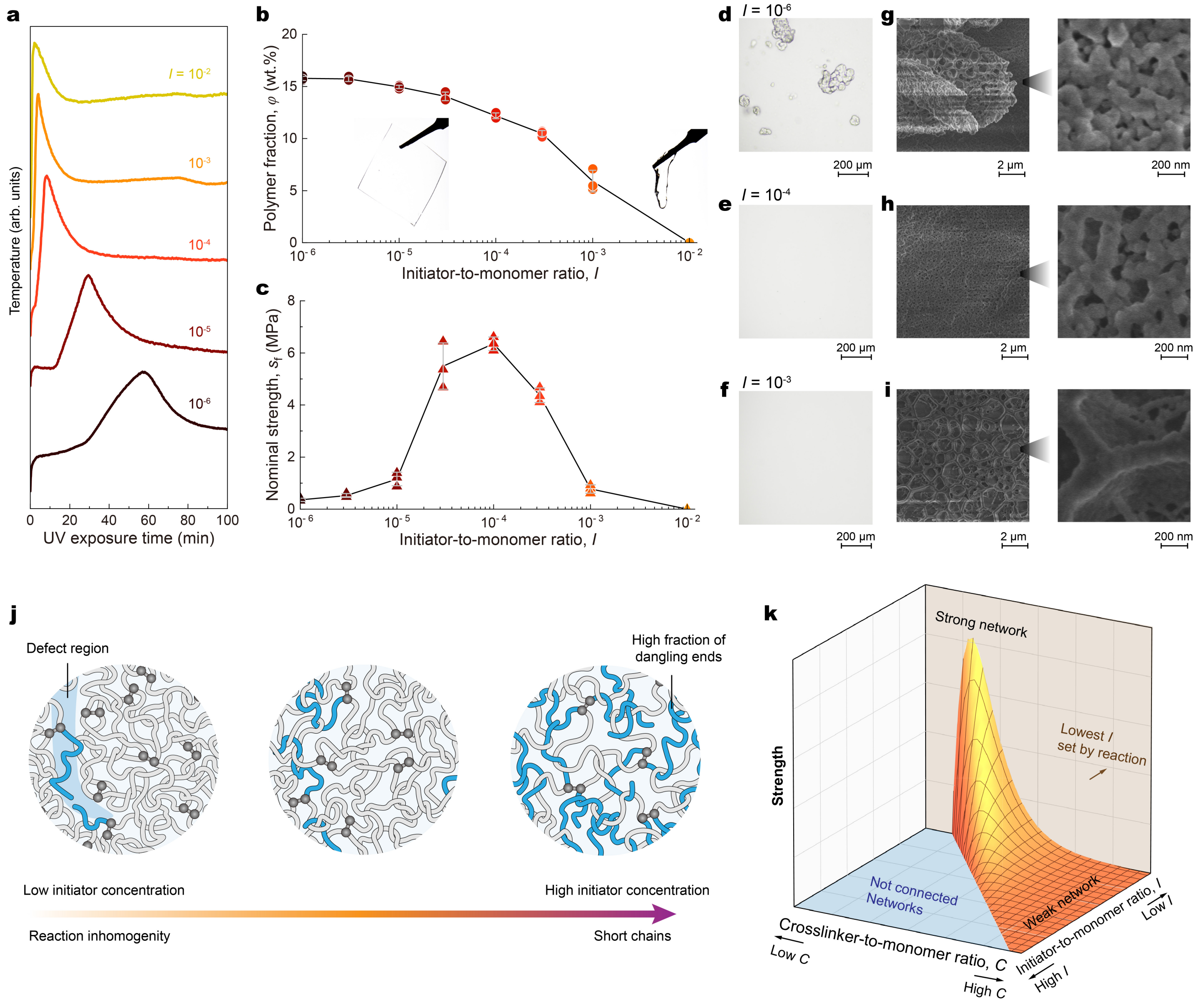

**Fig. 3. Competition between chain length and reaction homogeneity.** (a) Temperature of the reaction cell as a function of UV exposure time for different values of $I$. (b) Polymer fraction of the fully swollen hydrogels as a function of $I$. (c) Nominal strength of the fully swollen hydrogels as a function of $I$ . Image of the hydrogel under the microscope: (d) $I = 10^{-6}$, (e) $I = 10^{-4}$, and (f) $I = 10^{-3}$. Image of the cross-section of the hydrogel under the cryo-SEM: (g) $I = 10^{-6}$, (h) $I = 10^{-4}$, and (i) $I = 10^{-3}$. In all the experiments, we fix $C = 10^{-5}$. (j) Schematic of the network as a function of $I$. (k) Strength-synthesis relation. Data in (b) and (c) are presented as mean values ± standard deviations (s.d.) of $n = 3$ replicates.

To investigate this discrepancy between theoretical expectation and experimental observation, we image the hydrogels under an optical microscope (Figs. 3d-f, Supplementary Fig. 4). Prepared at an extremely low $I = 10^{-6}$, the hydrogel contains many defect regions, some with sizes over 100 μm (Fig. 3d). The defect regions spread over the thickness of the sample, as observed by a confocal microscope (Supplementary Fig. 5). Prepared at $I = 10^{-4}$ or $I = 10^{-3}$, the hydrogel has no such large defect regions (Figs. 3e-f).

We attribute the existence of optimal $I$ to two competing effects (Fig. 3j). At very low $I$, polymer chains grow very long, but polymerization proceeds inhomogeneously because the reaction is highly sensitive to oxygen and other impurities. As a result, certain regions remain unreacted, forming large defects. At very high $I$, polymerization remains homogeneous, but the polymer chains become short, leading to a high fraction of dangling ends. Therefore, for a given crosslink density, strength peaks at an intermediate $I$, where the two competing effects are balanced.

The above interpretation is further supported by the observation under cryo-scanning electron microscope (cryo-SEM) (Figs. 3g-i). We use this image to further assess spatial uniformity among the hydrogels. The hydrogel prepared at $I = 10^{-6}$ contains localized regions with substantially coarser pore morphology than the surrounding materials, consistent with the defect regions identified by optical microscopy (Fig. 3g). By comparison. At $I = 10^{-4}$, the observed morphology is fine and homogeneous (Fig. 3h). At $I = 10^{-3}$, the pore morphology is homogeneous but becomes coarse, which is attributed to its excessive swelling (Fig. 3i).

The minimum $I$ required for homogeneous polymerization is insensitive to $C$ (Supplementary Figs. 6-7). The polymerization kinetics is governed by radical generation and thus by the initiator concentration. The crosslinker is a species of monomer, which takes a small fraction and has little effect on the reaction rate (Supplementary Fig. 8). The minimum $I$ for homogeneous polymerization can be identified at a fixed $C$, and applies across a broad range of crosslink densities.

These considerations identify a synthesis-property relation in the space of the two synthesis variables ( $I$ and $C$), and strength (Fig. 3k). In this space, high strength occupies only a sharp peak, while the vast majority lies in a broad valley of low strength. This landscape helps explain why polyacrylamide hydrogels, despite decades of history and development, still remain weak (Fig. 1b, Supplementary Fig. 9). Conventional formulations largely fall within this broad valley. Our work illustrates a general concept of polymer science: synthesis creates topology of a network, and the topology determines properties. Here, guided by both the synthesis-topology relation and the topology-strength relation, we identify the sharp peak in a broad valley.

This synthesis-topology-property relation specifies how to achieve high strength within a given chemistry. The location of the high-strength region also depends on the monomer chemistry and polymerization condition. For a given $I$, the chain length also depends on the propagation and chain-transfer kinetics. For the photoinitiated polymerization, the light intensity also influences the chain length (Supplementary Fig. 10). Additionally, precursor purity becomes particularly important in preparing long chains. In common practice, the small amount of initiator is often introduced into the precursor as a dilute solution in alcohol. However, we find that even trace amounts of alcohol shorten the chains and reduce strength (Supplementary Fig. 11). We therefore avoid alcohol, but use water as the solvent for photoinitiator throughout this work in synthesizing hydrogels.

**Exceptional strength without compromising other properties**

Topology allows strength to be changed markedly without altering many other responses. To illustrate this point, we synthesize two networks that have similar stress-stretch curves before rupture (Fig. 4a), rheological behavior (Fig. 4b), and low hysteresis (Fig. 4c-d). Yet, the two networks differ in strength by over one order of magnitude (Fig. 4a).

This striking contrast exists because unlike viscoelastic behavior that is determined by the collective response of the majority of the network, rupture is an extreme and rare event that is dominated by localized high tension within a minority of the network. These considerations are supported by the markedly different stress-stretch curves in Fig. 4a. The weak network stiffens negligibly before rupture, because high tension concentrates in a small subset of strands, while the majority of strands remain coiled and underutilized. By contrast, the strong network stiffens steeply, because high tension spreads across many strands, which straighten simultaneously (Supplementary Note 3).

By making use of the potential of polymer chains to bear high tension, we achieve high strength without compromising other mechanical properties. For example, the high strength does not require building an energy dissipation mechanism, so the network maintains nearly perfect elasticity. Over a broad range of frequencies, the storage modulus exceeds the loss modulus by nearly two orders of magnitude (Fig. 4b). Under cyclic stretch to two times its original length, the loading and unloading curves nearly coincide (Fig. 4d). The hydrogel can be compressed to a small fraction of its initial height and bounce back elastically (Supplementary Video 2) In the plane of hysteresis and strength, the current work occupies the corner of the low hysteresis and high strength (Fig. 4f).

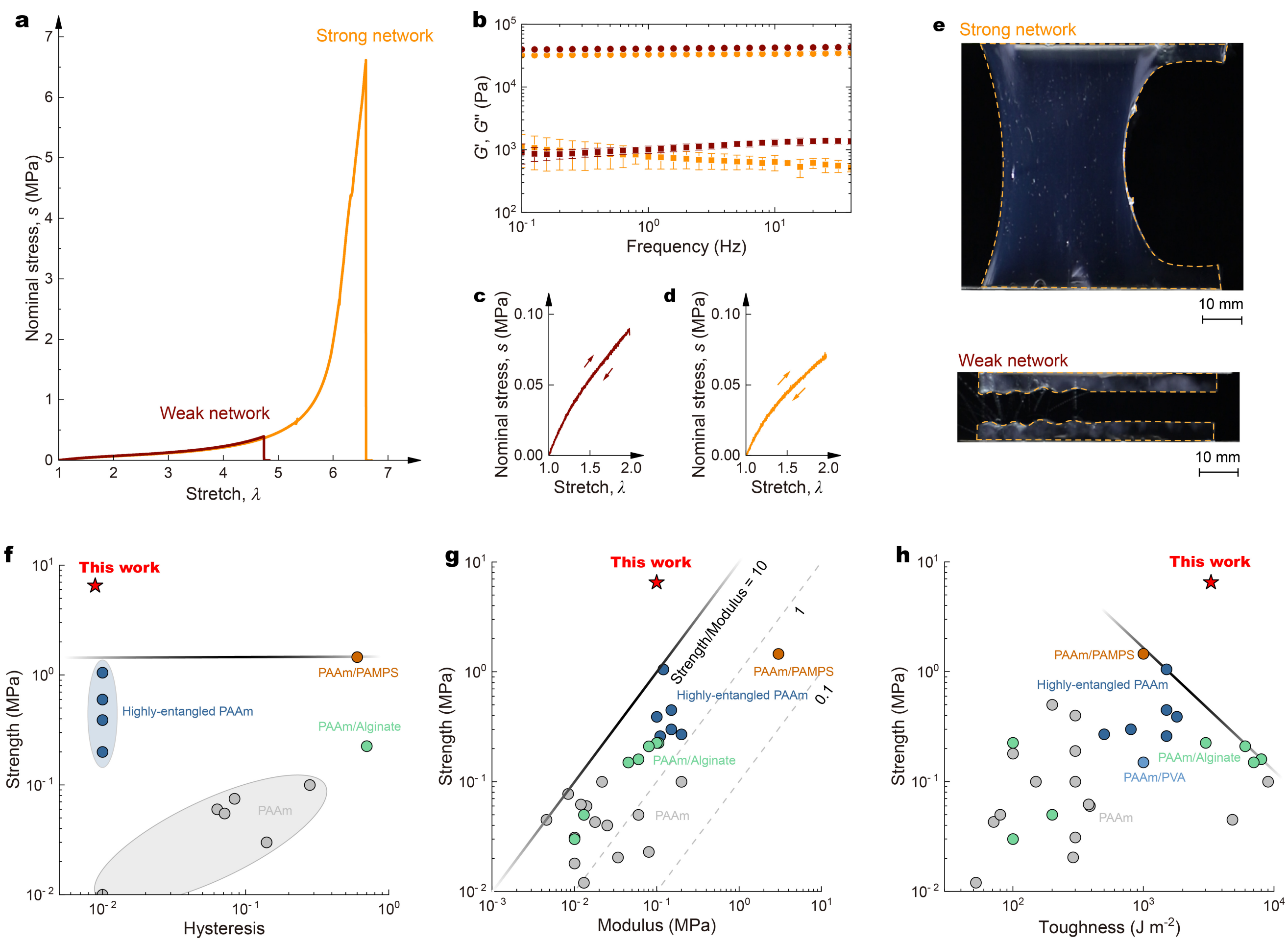

**Fig. 4. Exceptional strength without compromising other properties.** (a) Stress-stretch curve of a weak network of short chains and short strands ($C = 2\times10^{-4}$, $I = 10^{-2}$), and a strong network of long chains and long strands ($C = 10^{-5}$, $I = 10^{-4}$). (b) Rheological behaviors of the two networks. Loading and unloading curves of the (c) weak network, and (d) the strong network. (e) Stretch the precracked samples of the two networks. Materials on a plane of (f) hysteresis and strength, (g) modulus and strength, and (h) toughness and strength. Hysteresis is defined by the ratio of two areas: the area enclosed by the loading and unloading curves, and the area underneath the loading curve. All the hysteresis data are measured at a stretch of two. Data is collected from the literature[1,11-18,28-30].

Strength can also be decoupled from stiffness. In many polymer networks, high strength always accompanies high modulus (Fig. 4g). Here we show that a strong network can remain soft. Modulus at small deformation is governed by entropic elasticity of the average response of the network, whereas the strength at large deformation is limited by high tension of some minority of the network.

Strength is distinct from toughness. Toughness quantifies the resistance to crack growth, which has been greatly enhanced during the recent decade[19,20,27,31-34]. However, these efforts often do not enhance strength. For example, a PAAm hydrogel can have toughness approaching $10^5$ J $m^{-2}$,[14] but its strength still remains at 0.1 MPa (Fig. 4h). Since the work of Griffith[35], low strength has often been attributed to flaws that concentrate stress[36,37]. However, experimental evidence has accumulated that polymers can be insensitive to large flaws[1,4,14,38-40]. In fact, toughness and strength conflict for many classes of materials, including metals, ceramics, and plastics[41]. In this work, we show high strength and high toughness can be achieved simultaneously (Fig. 4e and h, Supplementary Video 3). The strong hydrogel has a toughness of ~3,300 J $m^{-2}$, measured by using the notched sample (Supplementary Fig. 12), and has a work of fracture of ~ 4 MJ $m^{-3}$, calculated by the area underneath the stress-stretch curve of the unnotched sample. The ratio of toughness to work of fracture defines a material-specific quantity with the dimension of length, called the fractocohesive length, which quantifies the flaw sensitivity of the material[38]. For the strong PAAm hydrogel, the fractocohesive length is ~0.8 mm, which is comparable to many other PAAm hydrogels[1,14,20]. Under the microscope, we do not observe flaws with size comparable to the fractocohesive length, indicating that rupture of the strong hydrogel is unlikely induced by the preexisting flaws.

## Apply the topological criterion to elastomers

Our topological criterion takes the advantage of invoking familiar topological features of a covalent network, such that it is readily applicable in elastomers and gels. As mentioned above, the lowest $I$ that produces long chains while maintaining the reaction homogeneity is nearly independent of $C$. These considerations allow us to amplify the strength following two steps. First, at a fixed crosslink density $C$, we find this lowest $I$ that maintains the reaction homogeneity. Second, using this optimal $I$ to find $C$ that balances network connectivity and the spread of high tension (Fig. 5a).

We apply this method to the synthesis of an elastomer, poly(ethyl acrylate) (PEA), a widely used polymer for coatings and adhesives (Figs. 5b-e). Both $I$ and $C$ again produce peak strength (Figs. 5b,c), consistent with the same competing effects identified in PAAm hydrogels. The optimized PEA network reaches a strength of approximately 10 MPa, about three times the values reported previously (Figs. 5d,e). Meanwhile, the network maintains low hysteresis (Supplementary Fig. 13).

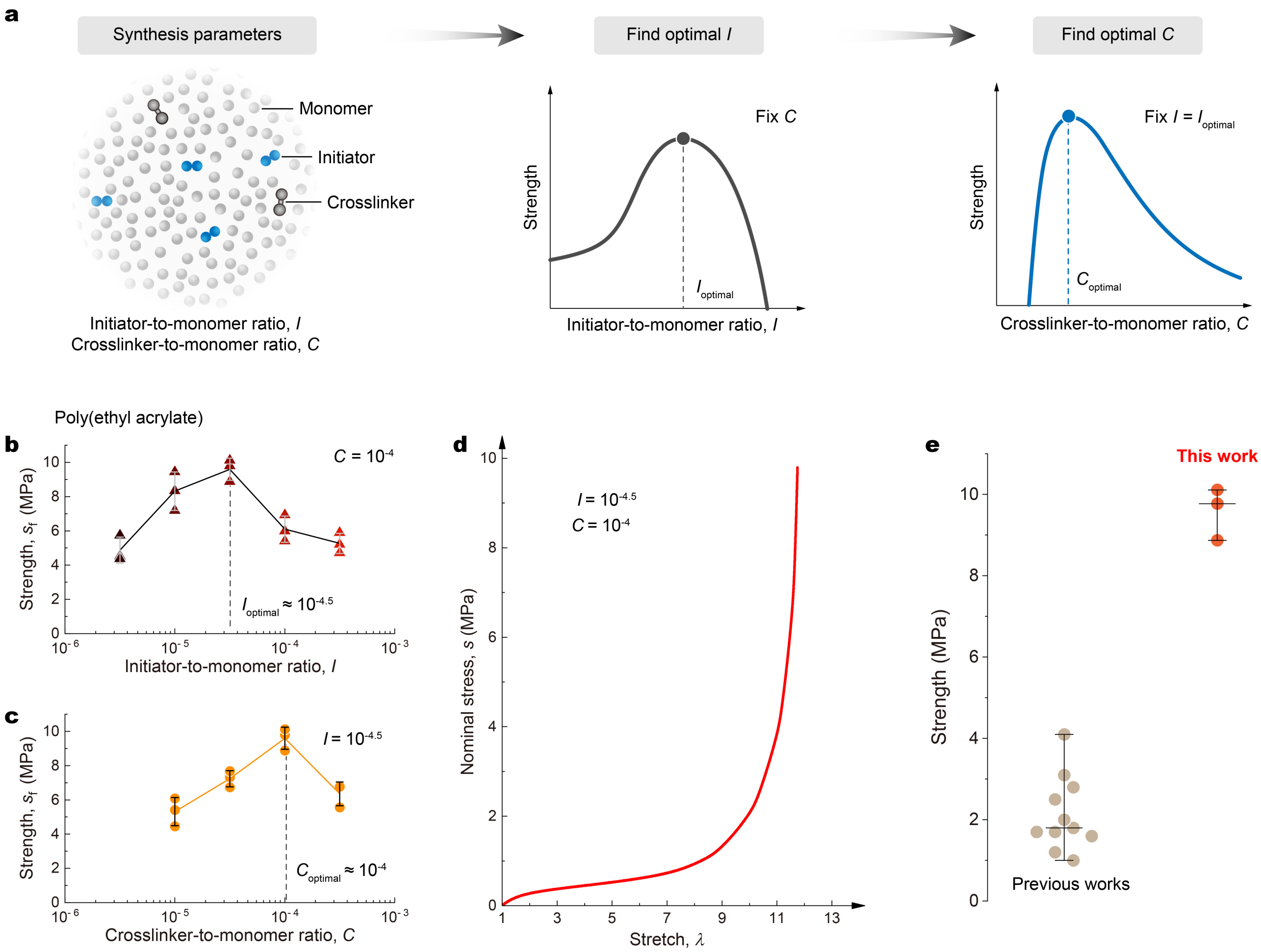

**Fig. 5. An elastomer with high strength.** (a) The strength is amplified by two steps. (b) For a poly(ethyl acrylate) (PEA) elastomer, at a fixed $C = 10^{-4}$, strength maximizes at an $I \approx 10^{-4.5}$. (c) At the optimal $I = 10^{-4.5}$, strength maximizes at $C = 10^{-4}$. (d) The stress-stretch curve of the PEA elastomer with optimal $I = 10^{-4.5}$, and $C = 10^{-4}$. (e) Strength of the PEA in this work compared with previous works[9,15,42,43]. Data in (b) and (c) are presented as mean values ± standard deviations (s.d.) of $n = 3$ replicates.

The topological criterion extends beyond specific chemistry. In many polymer networks, non-monotonic dependence of strength on crosslinks has long been documented[2,3,44,45]. Its physical origin, however, remained unresolved. The current work links the experimental observation to a physical origin, and to a topological criterion that balances conflicting effects between connectivity of the network, spread of high tension, and restriction of excessive stretchability.

## Concluding remarks

In summary, this work shows that the weakness of a polymer network arises not from individual covalent bonds, but from a topology that inefficiently recruits covalent bonds to bear high tension. We describe the topology of a network using five features and three independent variables. We identify how these topological features interplay to strengthen a network. The proposed topological criterion indicates that a network maximizes its strength through optimally sparse crosslinks, long chains, and dense entanglements. Each individual variable alone is insufficient for high strength, yet exceptional strength emerges when they are combined within a narrow regime. We find the regime by identifying a topology-strength relation and its physical origin. Accessible topologies are limited by synthesis. We identify a synthesis-topology relation that requires an optimal initiator concentration. Overly high initiator concentration reduces chain length, whereas overly low initiator concentration generates spatial heterogeneity, both reducing strength. Our work establishes a general synthesis–topology–property relation, suggesting that many polymer networks known to be weak can be strengthened through topology.

## Methods

### Materials

Acrylamide (AAm, A108470, Electrophoresis grade), *N*,*N*- Methylenebisacrylamide (MBAA, M128783, ≥ 99%), 2-Hydroxy-4′-(2-hydroxyethoxy)-2-methylpropiophenone (Irgacure 2959, H137984, ≥ 98.0%) are purchased from Aladdin Scientific Ltd. Ethyl acrylate (EA, E9706, 99%), tricyclo[5.2.1.0$^{2,6}$]decanedimethanol diacrylate (TDDA, 496669), and 2-hydroxy-2-methylpropiophenone (Irgacure 1173, 405655) are purchased from Sigma Aldrich. All the chemicals are used as purchased, without further purification. Deionized water is obtained from Mili-Q ultrapure water systems.

### Preparation of polyacrylamide (PAAm) hydrogels

The hydrogels are synthesized from precursors of four components: monomer (AAm), crosslinker (MBAA), photoinitiator (Irgacure 2959), and water. When the amount of crosslinker or photoinitiator is too small to weight precisely, we dissolve the crosslinker or photoinitator in water to form an aqueous solution with a concentration of 0.1 wt%, and use pipette to add the crosslinker or photoinitatior solution into the precursor. The four components are mixed in a centrifuge tube. The tube is sealed and soaked in a water bath of ~60 °C to facilitate the dissolution of AAm. The mixture is ultrasonicated for 3 minutes at room temperature to degas. A mold is prepared by attaching a hollow spacer of soft silicone rubber to a glass substrate.  The spacer has a thickness of 0.5 mm. The precursor solution is poured into the mold, which is subsequently covered with a glass sheet and securely clamped with binder clips. We expose the whole assembly to the ultraviolet (UV) light in a UV curing box, with an intensity of 6 mW cm$^{-2}$. To prevent the excessive chain branching reaction at the reaction stage of high monomer conversion, we monitor the temperature of the reaction cell throughout the polymerization, and stop the reaction after the temperature peak has passed. The reaction time differs for the precursor of different amounts of initiator (Fig. 3a). The sheet of the as-polymerized hydrogel is immersed in pure water to swell to equilibrium. We test all the properties at the fully swollen state.

### Preparation of poly(ethyl acrylate) (PEA) elastomers

The hydrogels are synthesized from precursors of three components: monomer (EA), crosslinker (TDDA), photoinitiator (Irgacure 1173). When the amount of crosslinker or photoinitiator is too small to weight precisely, we dissolve the crosslinker or photoinitator in EA to form a solution with a concentration of 1 wt%, and use pipette to add the crosslinker or photoinitatior solution into the precursor. The three components are mixed and sealed in a centrifuge tube, and are ultrasonicated for 3 minutes at room temperature to degas. A mold is prepared by attaching a hollow spacer of soft silicone rubber to a glass substrate. The spacer has a thickness of 0.8 mm. The other procedure is the same as in the preparation of PAAm hydrogels. The elastomer sheet is stored at a fume hood for 2 days to evaporate the residual monomers, after which we test all the properties.

Uniaxial tensile test

Nominal strength ($s_f$), stretchability ($\lambda_f$) are measured by monotonically loading a dogbone-shaped sample until rupture. The sample is cut from a sheet by a standard cutting die. The dogbone-shaped sample has a gauge region with a width of ~2 mm. The thickness of the sample is measured by a thickness meter. Two lines are marked on the gauge region of the sample at the undeformed state, with the distance between the two markers being the gauge length $L$. In the deformed state, the distance between the two markers becomes $l$. Stretch is calculated by $\lambda = l / L$. Nominal stress, $s$, is calculated by the force divided by the cross-sectional area of the undeformed sample. The stretch at rupture defines the stretchability ($\lambda_f$), and the nominal stress at rupture defines the nominal strength ($s_f$). Each sample is stretched with a fixed loading rate of 0.05 $s^{-1}$. Video of the test is recorded by a camera and the distance between mark lines is measured by ImageJ. During all tensile tests for hydrogels, a humidifier is used to prevent the hydrogel from losing water.

Cyclic tensile test is conducted at a fixed loading rate of 0.05 $s^{-1}$. Hysteresis is defined by the ratio of two areas: the area enclosed by the loading and unloading curves, and the area underneath the loading curve.

Toughness $G_c$ is measured by using two samples in a pure shear geometry. The sample is glued to rectangular acrylate sheets at both edges. A rectangular sample with the same width and height but without a precrack is stretched to obtain the stress-stretch curve, $s(\lambda)$. The elastic energy density is given by integrating the stress-stretch curve, $W(\lambda) = \int_1^\lambda s(\lambda)\mathrm{d}\lambda$. The energy release rate is given by the elastic energy density multiplying the height of the precacked sample, $G(\lambda) = W(\lambda)h_0$. The stretch rate is 0.05 $s^{-1}$. Toughness is defined as the energy release rate at which the crack irreversibly propagates through the entire sample, $G_\mathrm{c} = W(\lambda_\mathrm{c})h_0$.

Rheology test

Storage and loss moduli of the hydrogels are measured using a rheometer (Anton Paar, MCR 302) equipped with a 25 mm-diameter parallel-plate geometry. A normal preload of ~10 N is applied to ensure firm contact between the sample and the plates. Oscillatory shear measurements are performed at a fixed strain amplitude of 0.1%, with the frequency swept from 40 to 0.01 Hz at 25 °C.

Cryogenic scanning electron microscopy (Cryo-SEM)

A fully swollen hydrogel sample is mounted on a holder and quenched in liquid nitrogen. The sample, together with the holder, is transferred to a high-vacuum sputter coater (Leica EM ACE600), with both the transfer process and the coater maintained at −160 °C. Inside the coater, the sample is fractured to expose a cross-section. The temperature is then increased to −90 °C and held for 20 min to sublimate surface water. Subsequently, the sample is cooled back to −160 °C and sputter-coated with an approximately 5 nm-thick platinum layer. Finally, the sample is transferred to a cryogenic scanning electron microscope (Helios Nanolab G3 UC) for observation at −140 °C.

**Data availability**

The data in this study are available in the main article, supplementary information, and source data file. All data underlying this study are available from the corresponding author upon request

**Author Information**

These authors contributed equally: Zheqi Chen, Yu Zhou.

Contributions

Z.C. conceived the project and designed the study. Z.C. proposed the topological criterion, with the discussion and input from all authors. Z. C. and Y. Z. conducted the experiments. Z.C., Y.Z., C.L., H.C.S. and Z.S. discussed the results. Z.C., Y.Z., C.L., and Z.S. wrote the manuscript. Z. S. supervised the project.

Corresponding author

Correspondence to Zhigang Suo

**Supplementary Information**

Supplementary Notes 1-3, Supplementary Figs. 1-13, Captions for Supplementary Videos 1-3, additional references.

**Acknowledgements**

This work is supported by the National Science Foundation under MRSEC (DMR-2011754). Z.C. and H.C.S. acknowledge the support by the Health@InnoHK initiative of the Innovation and Technology Commission of the Hong Kong Special Administrative Region Government.

**Conflict of interest**

The authors declare no conflict of interest.

**References**

1. Yang, C., Yin, T. & Suo, Z. Polyacrylamide hydrogels. I. Network imperfection. *J. Mech. Phys. Solids*. **131**, 43-55 (2019).
2. Taylor, G. R. & Darin, S. R. The tensile strength of elastomers. *J. Polym. Sci.* **17**, 511-525 (1955).

3. Bueche, F. Tensile strength of rubbers. *J. Polym. Sci.* **24**, 189-200 (1957).
4. Zhao, X. Designing toughness and strength for soft materials. *Proceedings of the National Academy of Sciences*. **114**, 8138-8140 (2017).
5. Wang, S., Fan, Z., Gupta, C., Siavoshani, A. & Smith, T. Fracture Behavior of Polymers in Plastic and Elastomeric States. *Macromolecules*. **57**, 3875-3900 (2024).
6. Ito, K. Novel Cross-Linking Concept of Polymer Network: Synthesis, Structure, and Properties of Slide-Ring Gels with Freely Movable Junctions. *Polym. J.* **39**, 489-499 (2007).
7. Akagi, Y., Sakurai, H., Gong, J. P., Chung, U. & Sakai, T. Fracture energy of polymer gels with controlled network structures. *The Journal of Chemical Physics*. **139**, 144905 (2013).
8. Huang, J., Liu, J. & Lin, S. Topological mechanics of entangled networks. *ArXiv Preprint ArXiv:2509.17813*, (2025).
9. Zhou, Y., Li, C., Chen, Z., Bao, X. & Suo, Z. Elastic polymer networks of exceptional strength by deconcentrating tension. *ArXiv Preprint ArXiv:2610.11837*, (2026).
10. Chen, Z., Song, C., Shum, H. C. & Suo, Z. Heteropolymers of liquid-liquid phase separation achieve exceptional elasticity and strength., submitted (2026).
11. Wang, Y., Nian, G., Kim, J. & Suo, Z. Polyacrylamide hydrogels. VI. Synthesis-property relation. *J. Mech. Phys. Solids*. **170**, 105099 (2023).
12. Zhang, E., Bai, R., Morelle, X. P. & Suo, Z. Fatigue fracture of nearly elastic hydrogels. *Soft Matter*. **14**, 3563-3571 (2018).
13. Zheng, D., Lin, S., Ni, J. & Zhao, X. Fracture and fatigue of entangled and unentangled polymer networks. *Extreme Mech. Lett.* **51**, 101608 (2022).
14. Ma, J. et al. Designing Ultratough Single-Network Hydrogels with Centimeter-Scale Fractocohesive Lengths via Inelastic Crack Blunting. *Adv. Mater.* **36**, 2311795 (2024).
15. Kim, J., Zhang, G., Shi, M. & Suo, Z. Fracture, fatigue, and friction of polymers in which entanglements greatly outnumber cross-links. *Science*. **374**, 212-216 (2021).
16. Lei, J., Gao, Y., Xu, S., He, L. & Liu, Z. The effect of the effective polymer network on the extremely large deformation and fracture behaviors of polyacrylamide hydrogels. *J. Mech. Phys. Solids*. **200**, 106124 (2025).
17. Lee, S., Huang, Y., Dai, J., Kim, H. & Kim, J. Trapping Polymer Entanglements via Prolonged Initiation. *Macromolecules*. **59**, 5703-5710 (2026).
18. Guo, Y., Zhu, H., Chen, Z. & Qu, S. The mapping relationship between initiator and mechanical properties of free radical polymerized hydrogels. *Extreme Mech. Lett.* **77**, 102327 (2025).
19. Li, X. & Gong, J. P. Design principles for strong and tough hydrogels. *Nat. Rev. Mater.* **9**, 380-398 (2024).
20. Chen, Z. & Suo, Z. Thermodynamic and Molecular Origins of Crack Resistance in Polymer Networks. *Chem. Rev.* **126**, 606-670 (2026).
21. Mohanty, S., Blanchet, J., Suo, Z. & Cai, W. Why is the strength of an elastomeric polymer network so low. *Npj Comput. Mater.* **12**, 269 (2026).
22. Yu, Z. & Jackson, N. E. Shortest Paths Govern Bond Rupture in Thermoset Networks. *Macromolecules*. **58**, 1728-1736 (2025).
23. Furuya, T. & Koga, T. Molecular Simulation of Effects of Network Structure on Fracture Behavior of Gels Synthesized by Radical Polymerization. *Macromolecules*. **58**, 3359-3368 (2025).
24. Li, J. et al. Inherent Tensile Strength and Stretchability of Unentangled Elastomers. *Macromolecules*. **58**, 11445-11464 (2025).
25. Granick, S. & Rubinstein, M. A multitude of macromolecules. *Nat. Mater.* **3**, 586-587 (2004).
26. Rubinstein, M. & Panyukov, S. Elasticity of Polymer Networks. *Macromolecules*. **35**, 6670-6686 (2002).
27. Panyukov, S. Theory of Flexible Polymer Networks: Elasticity and Heterogeneities. *Polymers*.

**12**, 767 (2020).
28. Bai, R., Yang, J., Morelle, X. P. & Suo, Z. Flaw-Insensitive Hydrogels under Static and Cyclic Loads. *Macromol. Rapid Commun.* **40**, 1800883 (2019).
29. Sun, J. et al. Highly stretchable and tough hydrogels. *Nature*. **489**, 133-136 (2012).
30. Gong, J. P. Why are double network hydrogels so tough? *Soft Matter*. **6**, 2583-2590 (2010).
31. Creton, C. 50th Anniversary Perspective: Networks and Gels: Soft but Dynamic and Tough. *Macromolecules*. **50**, 8297-8316 (2017).
32. Wang, M., Hu, J. & Dickey, M. D. Tough Ionogels: Synthesis, Toughening Mechanisms, and Mechanical—A Perspective. *JACS Au*. **2**, 2645-2657 (2022).
33. Xu, C. et al. Mechanical Regulation of Polymer Gels. *Chem. Rev.* **124**, 10435-10508 (2024).
34. Cho, Y. E., Lee, S., Ma, S. J. & Sun, J. Network design for soft materials: addressing elasticity and fracture resistance challenges. *Soft Matter*. **21**, 1603-1623 (2025).
35. Griffith, A. A. VI. The phenomena of rupture and flow in solids. *Philosophical Transactions of the Royal Society of London, Series a: Containing Papers of a Mathematical Or Physical Character*. **221**, 163-198 (1921).
36. Lake, G. J. & Thomas, A. G. The Strength of Highly Elastic Materials. *Proceedings of the Royal Society of London. Series a, Mathematical and Physical Sciences*. **300**, 108-119 (1967).
37. Gent, A. N. 10 - Strength of Elastomers. In: Mark, J. E., Erman, B. & Eirich, F. R., editors. *Science and Technology of Rubber (Third Edition)*. Burlington: Academic Press; 2005. pp. 455-495.
38. Chen, C., Wang, Z. & Suo, Z. Flaw sensitivity of highly stretchable materials. *Extreme Mech. Lett.* **10**, 50-57 (2017).
39. Long, R., Hui, C., Gong, J. P. & Bouchbinder, E. The Fracture of Highly Deformable Soft Materials: A Tale of Two Length Scales. *Annual Review of Condensed Matter Physics*. **12**, 71-94 (2021).
40. Koshut, W. J. et al. Flaw sensitivity and tensile fatigue of a high-strength hydrogel. *Int. J. Fatigue*. **163**, 107071 (2022).
41. Ritchie, R. O. The conflicts between strength and toughness. *Nat. Mater.* **10**, 817-822 (2011).
42. Steck, J., Kim, J., Kutsovsky, Y. & Suo, Z. Multiscale stress deconcentration amplifies fatigue resistance of rubber. *Nature*. **624**, 303-308 (2023).
43. Zhang, G., Steck, J., Kim, J., Ahn, C. H. & Suo, Z. Hydrogels of arrested phase separation simultaneously achieve high strength and low hysteresis. *Sci. Adv.* **9**, eadh7742 (2023).
44. Bueche, F. & Dudek, T. J. Tensile Strength of Amorphous Gum Rubbers. *Rubber Chem. Technol.* **36**, 1-10 (1963).
45. Flory, P. J., Rabjohn, N. & Shaffer, M. C. Dependence of tensile strength of vulcanized rubber on degree of cross-linking. *J. Polym. Sci.* **4**, 435-455 (1949).